# Current-voltage characteristic of parallel-plane ionization chamber with inhomogeneous ionization

**Dimitar G. Stoyanov**

Faculty of Engineering and Pedagogy in Sliven, Technical University of Sofia
59, Bourgasko Shaussee Blvd, 8800 Sliven, BULGARIA

E-mail: dgstoyanov@abv.bg

**Abstract:** The balances of particles and charges in the volume of parallel-plane ionization chamber are considered. Differential equations describing the distribution of current densities in the chamber volume are obtained. As a result of the differential equations solution an analytical form of the current-voltage characteristic of parallel-plane ionization chamber with inhomogeneous ionization in the volume is got.

## 1. Introduction

The ionization chamber is a simple and reliable device for measuring the characteristics of beams of particles and radiations. It concerns beams of radioactive radiation as well as X-rays, synchrotron radiation, beams of electrons, protons and neutrons generated in various accelerators.

The characteristics of these beams are measured by registering their interaction with the gas medium causing a partial ionization of the molecules of the medium. The generated charged particles are put under the action of electric field germinated by the potential difference between two metallic electrodes. That causes the current running through gas volume which current carriers are the ionization products.

As it is shown in figure 1 generally the beam of radiation passes through the gas volume of the ionization chamber so that the ionization takes place merely in a part of the ionization chamber volume. This imposes the modification of the model represented in [1] taking in account the ionization inhomogeneity in the gas volume of the ionization chamber. In simplicity we will consider a parallel-plane chamber. Besides, we will suppose that the beam embraces the whole chamber by height.

The objective of this article is the obtaining of an analytical form of current-voltage characteristic of parallel-plane chamber with inhomogeneous ionization in the volume.

## 2. Geometry and basic equations

We will consider a parallel-plane type of an ionization chamber. In this case both metallic electrodes account for parallel metallic planes disposed at a distance **d** each other (figure 1). The plates have surface **S** ( $\mathbf{d^2 \ll S}$ ). We assume that the electric field is concentrated only in the volume between the plates and it is homogeneous.

For the description of the spatial coordinates we choose a coordinate system with OX-axis perpendicular to the plates and axes OY and OZ making a plane parallel to the plates. Let the cathode and the anode have a coordinate $\mathbf{x_k = 0}$ and $\mathbf{x_a = d}$, respectively.

At such configuration of the electrodes the vectors of electric field strength and the velocities of movements of the charged particles will have directions parallel to the OX-axis. The vector of the electric field strength $\vec{\mathbf{E}}$ is pointed from the anode to the cathode.

In the making of the model of a parallel-plane ionization chamber with inhomogeneous ionization in the volume we will assume just like in [1] that the basic mechanism for the generation of charged particles in the volume is the ionization of gas molecules. Thus the generated charged particles could be lost mainly by two mechanisms [1, 2]:
- Neutralization of charged particles on the metallic electrodes;
- Volume recombination during the interaction between a positive ion and an electron in the volume in which they are neutralized.

The balances of the particles are written in the same manner as in [1] in a stationary case without reading the diffusion of the charged particles, and without reading the influence of the charged particles upon the electric field created by the metallic electrodes.

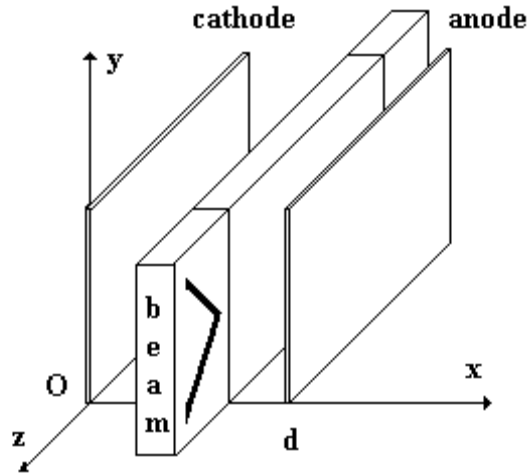

Figure 1. Model spatial location of a beam and parallel-plane chamber.

The electrons move in the direction from the cathode to the anode and they are absorbed by the anode. They form the electronic component of current density $\mathbf{j_e(x)}$ through the chamber.

The positive ions move in the direction from the anode to the cathode, and they are neutralized by the cathode. They form the ionic component of current density $\mathbf{j_+(x)}$ through the chamber.

The density of current through the chamber $\mathbf{j}$ is the sum of the electronic and ionic components:

$$\mathbf{j} = \mathbf{j_e} + \mathbf{j_+} \quad (1)$$

In the parallel-plane stationary case $\mathbf{j}$ is a constant in the gas volume.

If we express the concentrations of the charged particles through their currents $\mathbf{j_e(x)}$ and $\mathbf{j_+(x)}$, and mobilities $\mu_e$ and $\mu_+$, respectively, from the balance of the charged particles concentrations we obtain a differential equation which submits to the current density of the electrons $\mathbf{j_e(x)}$ [1]:

$$\frac{dj_e}{dx} = e.I_i(x) - \frac{\beta}{e.\mu_e.\mu_+.E^2} \cdot j_e \cdot (j - j_e) \quad (2)$$

where $\beta$ is a coefficient of two-particle recombination;

$\mathbf{e}$ is the electric charge of an electron;

$\mathbf{I_i(x)}$ is the ionization rate in a unit of volume per time unit.

This is the equation which solution we will examine further on.

During the solution of the equation the following boundary conditions upon the electrodes are taken into account [1, 2]:

- the cathode does not emit electrons [2]:

$$j_e(x=0) = 0 \qquad (3)$$

- the anode does absorb the electrons falling on it [2]:

$$j_e(x=d) = j \qquad (4)$$

Here, in this work on the contrary to the considered case in [1], where the ionization takes place in the whole volume, it is supposed that the rate of ionization in a unit of volume $I_i(x)$ is different in the different sections of the chamber volume. The reason for this is the presence of a beam of ionization radiation passing through the chamber volume without crossing the electrodes (see Figure 1) and without filling up the chamber volume.

Figure 2 represents the model dependence of $I_i(x)$, which we will use for the solution of (2) in this work

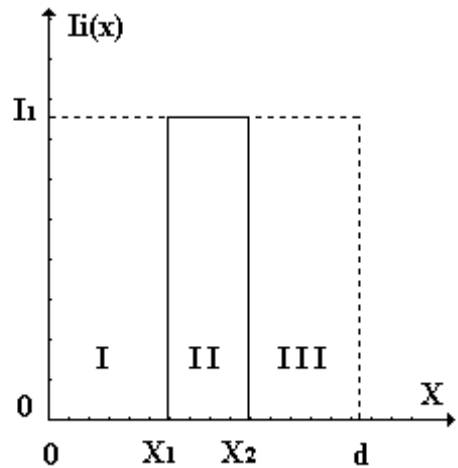

Figure 2. A distribution of ionization rate in a unit of volume.

Across the gas volume points with coordinates $x \in [0, x_1]$ the beam of radiation does not pass, because of that no ionization takes place (in fact the ionization has a zero value). This interval of coordinates we will term **Region I**.

In the gas volume points with coordinates $x \in [x_1, x_2]$ a beam of radiation passes and because of that ionization takes place. The ionization $I_i(x)$ is constant and its value is $I_1$. This interval of coordinates we will term **Region II.**

In the gas volume points with coordinates $x \in [x_2, d]$ no ionization takes place. This interval of coordinates we will term **Region III.**

### 3. Equation solution

In order to simplify the record when we solve (2) in the common case we make it dimensionless to $j_e$.

$$f(x) = \frac{j_e}{j} \in [0,1], \qquad (5)$$

The function $f(x)$ shows how the electronic current component in the vessel volume changes at a definite working regime. The function $f(x)$ accepts the boundary values $f(x = 0) = 0$ and $f(x = d) = 1$ on the cathode and anode, respectively.

Besides, this $1-f(x)$ will give us information how does the ionic component change in the volume.

Using (5) equation (2) could be transformed in

$$\frac{df}{dx} = \frac{e.I_i(x)}{j} - \frac{\beta.j}{e.\mu_e.\mu_+.E^2}.f.(1-f) \tag{6}$$

### 3.1 A solution in Region I
In Region I there is no ionization and the particles recombination could be entirely manifested.

In this region the differential equation gets the type

$$\frac{df}{dx} = -\frac{\beta.j}{e.\mu_e.\mu_+.E^2}.f.(1-f) = -b.f.(1-f) \tag{7}$$

where - $b = \dfrac{\beta.j}{e.\mu_e.\mu_+.E^2}$ \hfill (8)

The solving of (7) according to [3] gives the following common solutions:

$$f(x) = 0. \tag{9a}$$

$$f(x) = \frac{1}{1 + \exp[b.(x+C)]}. \tag{9b}$$

$$f(x) = 1. \tag{9c}$$

From these solutions only (9a) satisfies the boundary condition on the cathode. It follows from this that in this region there are only ions but electrons should not be available. This is the reason the recombination rate to be zero. Besides as a consequence is got the boundary condition

$$f(x_1) = 0. \tag{10}$$

### 3.2 A solution in Region III
In Region III there is no ionization and the particles recombination could be entirely manifested.

In this region the differential equation gets the type (7). From the solutions of (7), i.e. (9a), (9b) and (9c), only (9c) satisfies the boundary condition on the anode. It follows from this that in this region there are only electrons but ions should not be available. This is the reason the recombination rate to be zero. Besides as a consequence is got the boundary condition

$$f(x_2) = 1. \tag{11}$$

### 3.3 A solution in Region II
In Region II there is ionization and the particles recombination could be also manifested.

In this region the differential equation gets the type

$$\frac{df}{dx} = \frac{e.I_1}{j} - \frac{\beta.j}{e.\mu_e.\mu_+.E^2}.f.(1-f) = a + b.\left(f - \frac{1}{2}\right)^2 \qquad (12)$$

where - $a = \dfrac{e.I_1}{j} - \dfrac{1}{4}\dfrac{\beta.j}{e.\mu_e.\mu_+.E^2}$ \qquad (13)

### 3.3.1 *Regime: No Recombination*
We consider this uttermost case when either there is no recombination or the combination is slightingly small, e.g. at very stronght electric field.

In this case on the right side of equation (12) only the first member remains which does not depend on **x.**

Taking into account the boundary condition (10) the solution is very simple

$$f(x) = \frac{e.I_1}{j}.(x - x_1) \qquad (14)$$

For the fulfillment of the boundary condition (11) is necessary

$$j = e.I_1.(x_2 - x_1) = e.I_1.\delta = j_s. \qquad (15)$$

Here with $\delta = x_2 - x_1$ the thickness of the beam of radiation is marked.

When the chamber is in this working regime all charges generated in the chamber volume will be directed to the electrodes and they will reach them. That is the reason the current that is available at this regime to be maximum, and we will call it a *current of saturation* $j_s$.

### 3.3.2 *Regime: With recombination*
The solving of (12) according [3] gives the following common solution

$$\frac{1}{\sqrt{a.b}}.arctg\left[\sqrt{\frac{b}{a}}.\left(f - \frac{1}{2}\right)\right] = x + C \qquad (16)$$

After taking into account the boundary conditions (10) and (11) and some transformations the following is obtained

$$f(x) = \frac{1}{2} + \sqrt{\frac{a}{b}}.tg\left[\sqrt{a.b}.\left(x - x_1 - \frac{\delta}{2}\right)\right] \qquad (17)$$

In order the boundary conditions to be satisfied is necessary and sufficiently

$$\sqrt{\frac{a}{b}}.tg\left[\sqrt{a.b}.\left(\frac{\delta}{2}\right)\right] = \frac{1}{2} \qquad (18)$$

This formula is a transcendent equation for the relation between **a** and **b**, but in this case it plays the role of an analytical form of the current-voltage characteristic of the ionization chamber.

### 4. An analysis of the solution
In order to analyze the solution obtained we put

$$4.E_1^2 = \frac{\beta.\delta.j_s}{e.\mu_e.\mu_+} \quad (19)$$

and using (19) we could set the constants **a** and **b** of (12) in the form

$$a = \frac{j_s}{j.\delta} - \frac{E_1^2}{E^2}.\frac{j}{j_s.\delta} \quad (20)$$

$$b = 4.\frac{E_1^2}{E^2}.\frac{j}{j_s.\delta} \quad (21)$$

The current-voltage characteristic of the ionization chamber is hidden in (18). If we replace the constants with their equals from (20) and (21) and after some transformations we get

$$\sqrt{\frac{\frac{E^2}{E_1^2} - \frac{j^2}{j_s^2}}{\frac{j^2}{j_s^2}}}.tg\left[\frac{\sqrt{\frac{E^2}{E_1^2} - \frac{j^2}{j_s^2}}}{\frac{E^2}{E_1^2}}\right] = 1 \quad (22)$$

The equation (22) is transcendent and is solved numerically. The curve of the dependence between the current density and the electric field strength (the potential difference between both electrodes, respectively) is represented in Figure 3.

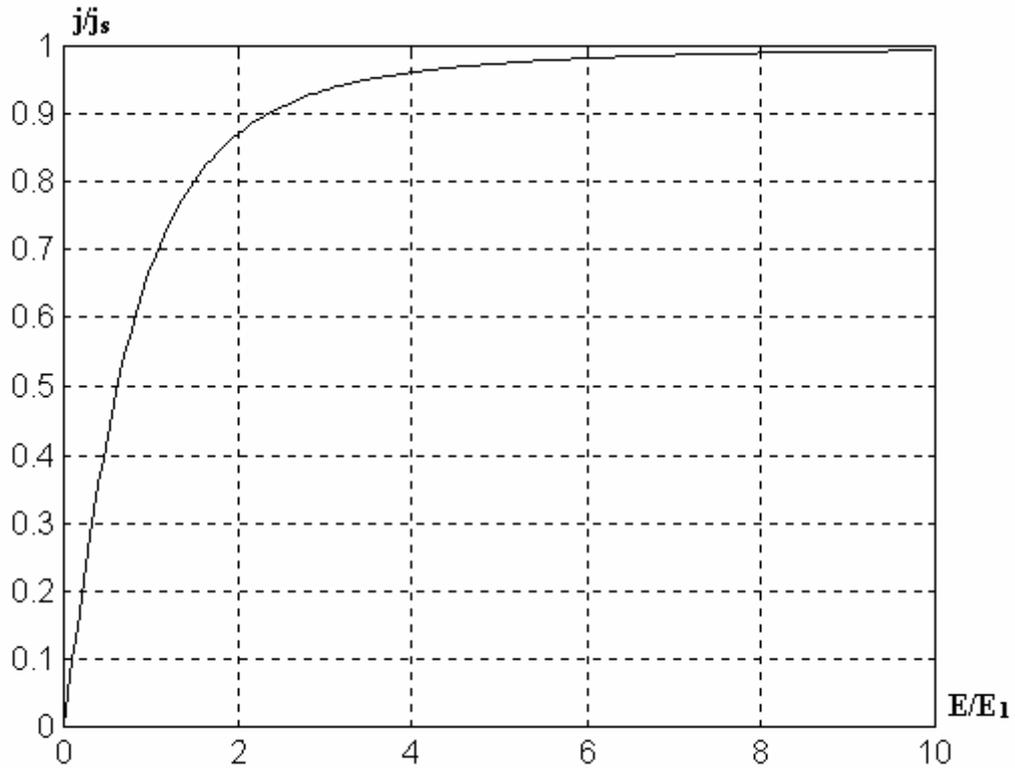

Figure 3. A current-voltage characteristic of parallel-plane ionization chamber with inhomogeneous ionization.

Here this is worthwhile is significant to be noted that the current-voltage characteristic of the ionization chamber with inhomogeneous ionization in the volume does not depend on the position of beam of radiation, but merely on its thickness.

Thus the obtained dependence (22) is similar in form to that got in [1] for the case of parallel-plane ionization chamber with homogeneous ionization in the chamber volume.

This the reason all comments about (22) represented in [1] to be also valid for this case. Just the same the approximation dependences look at strong and weak fields.

At strong fields the current-voltage characteristic could be approximately determined from (22) as

$$\frac{j}{j_s} \cong 1 - \frac{2}{3}\cdot\left(\frac{E_1}{E}\right)^2 + \frac{4}{5}\cdot\left(\frac{E_1}{E}\right)^4. \qquad (23)$$

At weak fields the recombination is strong, and in this case the current-voltage characteristic could be approximately determined from (22) as

$$\frac{j}{j_s} \cong \frac{E}{E_1}\cdot\left[1 - \frac{\pi^2}{8}\cdot\left(\frac{E}{E_1}\right)^2\right]. \qquad (24)$$

The differences are in the different values of $j_s$ and $E_1$, and besides that in the different dependences of these magnitudes on the parameters of the ionization chamber.

$$j_s = e.I_1.\delta = e.I_1.d.\frac{\delta}{d} = j_s(d).\frac{\delta}{d}. \qquad (23)$$

$$E_1 = \sqrt{\frac{\beta.\delta.j_s}{4.e.\mu_e.\mu_+}} = \sqrt{\frac{\beta.I_1.\delta^2}{4.\mu_e.\mu_+}} = \sqrt{\frac{\beta.I_1.d^2}{4.\mu_e.\mu_+}}\cdot\frac{\delta}{d} = E_1(d).\frac{\delta}{d}. \qquad (24)$$

New magnitudes $j_s(d)$ and $E_1(d)$ are introduced here, which account for a current of saturation and a characterizing electric field strength of the ionization chamber, in which the ionization is running in the whole chamber volume at one and the same $I_1$. Namely these magnitudes figure as $j_s$ and $E_1$ in [1].

Therefore, we may confirm that the present result and the one in [1] do not contradict but complement one another.

As an illustration in Figure 4 the dependence of current $j/j_s(d)$ across the chamber as a function of the thickness of the beam of radiation at set of field intensities $E/E_1(d)$ =0.25, 0.5, 1.0, 2., 4., 8.

As it is evident in the Figure 4 the current of saturation proportionally decreases with the decreasing of beam thickness. The saturation itself happens at lower strength values of the applied electric field.

**5 Conclusion**

In conclusion it might be said that are obtained differential equations which the current in the volume of a parallel-plane ionization chamber with inhomogeneous ionization submits to. An analytical form of the current-voltage characteristic of a parallel-plane ionization chamber with inhomogeneous ionization is got in the form of a transcendent equation.

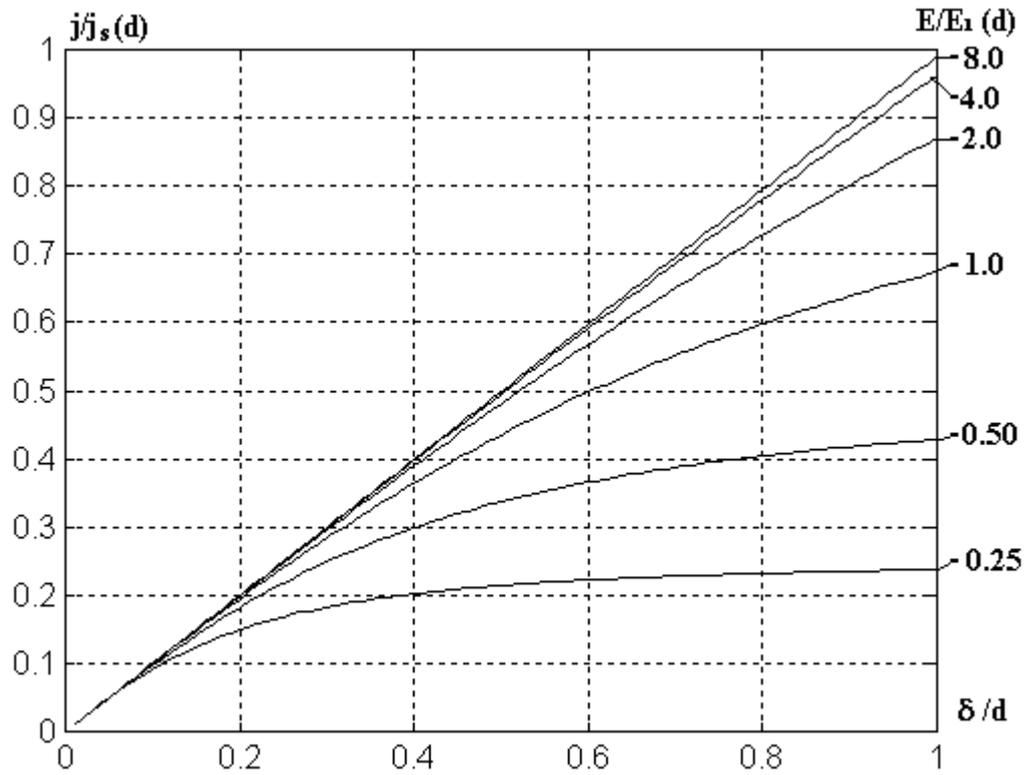

Figure 4. Dependence of the current across the chamber vs. the beam thickness